\documentclass[a4paper,11pt]{article}
\usepackage{float}
\usepackage[pdftex]{graphicx}
\usepackage{subcaption}
\usepackage[T1]{fontenc}
\usepackage{lmodern}
\usepackage{slashed}
\usepackage[utf8]{inputenc}
\usepackage[english]{babel}
\usepackage{microtype}
\usepackage{cite}
\usepackage{amsmath,amssymb,amsfonts,amsthm}
\usepackage{mathtools,mathrsfs,calligra,aurical}
\usepackage[nottoc,notlot,notlof]{tocbibind}
\usepackage{upgreek}
\usepackage{mathtools}
\numberwithin{equation}{section}
\allowdisplaybreaks
\usepackage[all]{xy}
\usepackage{color} 
\usepackage{xcolor}
\usepackage{graphicx}
\usepackage{cancel}
\graphicspath{{images/}}
\usepackage{geometry}
\usepackage[toc,page]{appendix}
\usepackage{hyperref}
\usepackage[normalem]{ulem}

\usepackage{bm}
\usepackage{ragged2e}
\usepackage{appendix}
\usepackage{slashed}
\usepackage{bbold}
\usepackage{cancel}

\definecolor{blue-violet}{rgb}{0.54, 0.17, 0.89}
\definecolor{PineGreen}{cmyk}{0.92, 0, 0.59, 0.25}
\definecolor{YellowOrange}{cmyk}{0, 0.42, 1, 0}
\definecolor{orange}{rgb}{0.95, 0.5, 0.1}

\newcommand{\be}{\begin{equation}}
\newcommand{\bea}{\begin{eqnarray}}

\newcommand{\ee}{\end{equation}}
\newcommand{\eea}{\end{eqnarray}}
\def\a{\alpha}\def\d{\partial}

\DeclareMathAlphabet{\mathpzc}{OT1}{pzc}{m}{it}

\usepackage[dvipsnames]{xcolor}

\begin{document}

\begin{titlepage}
\begin{flushright}
\par\end{flushright}
\vskip 0.5cm
\begin{center}
\textbf{\LARGE \bf  Boundary-Boundary Duality on Regular Black Holes, Supersymmetric Solitons and Holographic Spinning Plasma Disks}\\
\vskip 5mm

\vskip 1cm

\large {\bf Andr\'{e}s Anabal\'{o}n}$^{~a ~b}$\footnote{anabalo@gmail.com}  \large {and \bf Horatiu Nastase}$^{~b}$\footnote{horatiu.nastase@unesp.br}

\vskip .5cm 

$^{(a)}${\textit{Departamento de F\'isica, Universidad de Concepci\'on, Casilla 160-C, Concepci\'on, Chile.}}\\ \vskip .1cm
$^{(b)}${\textit{Instituto de F\'isica Te\'orica, UNESP-Universidade Estadual Paulista \\
R. Dr. Bento T. Ferraz 271, Bl. II, Sao Paulo 01140-070, SP, Brazil.}}
\end{center}
\begin{abstract}
We construct a new family of exact rotating solutions of four-dimensional Einstein--Maxwell theory with negative cosmological constant describing regular spinning AdS black holes and smooth solitons. The geometries are free of curvature singularities and admit open regions of parameter space without closed timelike curves, while supersymmetric limits correspond to causally regular horizonless configurations. A remarkable feature of these solutions is the existence of at least two codimension-one conformal boundaries. The first is Minkowski spacetime, where the expectation value of the energy--momentum tensor describes a finite spinning disk of strongly coupled conformal plasma whose edge rotates at the speed of light. The second boundary is a rotating black-hole geometry with vanishing energy--momentum tensor but non-vanishing Cotton tensor, defining a holographic theory in which the dual graviton is fixed at the boundary. We show that the energy density associated with the dual graviton exactly reproduces that of the plasma after a suitable analytic continuation and the identification of the holographic energy scale with the Lorentz factor of the rotating fluid, providing evidence for a boundary-boundary duality. Indeed, electromagnetic duality exchanges the electric and magnetic currents between the boundary supporting the graviton and that supporting the dual graviton thus providing a generalized mirror symmetry between these boundary theories. Hence, the Lorentz factor is identified with the electromagnetic dual of the renormalization scale. In an appropriate limit, the solutions reduce to the planar Reissner--Nordström--AdS black hole, whereas the supersymmetric solitons have no regular static limit.
\end{abstract}
\vfill{}
\vspace{1.5cm}
\end{titlepage}

\setcounter{footnote}{0}
\tableofcontents

\section{Introduction and Discussion}
The discovery that the quark--gluon plasma (QGP) produced in non-central heavy-ion collisions carries enormous vorticity has opened a new frontier in the study of strongly interacting matter. Measurements of the global polarization of $\Lambda$ hyperons at RHIC indicate that the QGP is the most vortical fluid ever observed, with characteristic vorticities reaching $\sim 10^{21}\,\mathrm{s}^{-1}$ for cells of characteristic length of $R\sim 10^{-15}\, \mathrm{m}$  \cite{STAR:2017ckg}. Remarkably, during the deconfined stage preceding hadronization, the QGP behaves as an almost perfect relativistic fluid, exhibiting an extraordinarily small viscosity-to-entropy ratio, near the holographic bound \cite{Policastro:2001yc}. Yet, despite the wealth of experimental data and the success of hydrodynamics in describing the bulk evolution of the plasma, our understanding of strongly coupled rotating fluids remains extremely limited. Unlike weakly coupled systems, where rotation can often be studied directly from microscopic degrees of freedom, there is currently no first-principles framework capable of describing the equilibrium properties of a generic strongly interacting fluid at large vorticity \cite{Becattini:2020ngo}. Exact equilibrium configurations of strongly coupled rotating relativistic fluids provide valuable laboratories in which these questions can be addressed non-perturbatively, helping to isolate universal consequences of strong coupling from the complex dynamical environment of heavy-ion collisions.

Holography \cite{Maldacena:1997re, Witten:1998qj} is a technique that allows to model this problem at strong coupling. In this context, a holographic description of a plasma ball is given by a solution to the Einstein--Maxwell--AdS sector of maximal supergravity with the boundary condition that the conformal boundary admits a metric with a representative of the form\footnote{Here, we focus our discussion on three-dimensional fluids, as this captures the essence of the problem, which is cylindrically symmetric at equilibrium in four dimensions. This is also the natural setting for AdS/CMT, and these results therefore provide an arena in which condensed matter phenomena can be analyzed in the presence of rotation. We adopt the convention that Greek indices run over three values for three-dimensional quantities and over four values for four-dimensional quantities. We hope this convention makes the notation less cluttered.}
\begin{equation} \label{b1}
g^{r}_{\mu \nu}dz^{\mu}dz^{\nu}=-dt^2+dw^2+w^2d\phi^2 \, .
\end{equation}
At the same time, the gravitational description must yield a VEV for the QFT energy-momentum tensor of the form of a conformal spinning fluid, with angular velocity $\Omega_0$ which is completely kinematically constrained up to an overall constant $E$,
\begin{equation} \label{EMtensor}
\langle T_{\mu \nu} \rangle
=
U_{\mu}U_{\nu}(\rho+P)+g^{r}_{\mu \nu}P \, ,
\end{equation}
with
\begin{equation}
P=\frac{\rho}{2}= \frac{E}{2} \gamma^3\, , \qquad
U=\gamma(\partial_{t}+\Omega_0\partial_{\phi})\, , \qquad
U^{\mu}U_{\mu}=-1\, , \qquad
\gamma=(1-w^2\Omega_0^2)^{-1/2} \, .
\end{equation}
An educated guess, guided by these geometrical and dynamical considerations, led us to the exact solution presented in \eqref{metric}, together with the gauge field \eqref{A}\footnote{When there is a cosmological constant, there is no known method to generate solutions to the Einstein equations nor to the Einstein-Maxwell equations.}. We show below that the explicit calculation of the holographic energy-momentum tensor yields
\begin{equation}
E=\frac{L^4}{\kappa}m\Omega_0^3 \, .
\end{equation}
$m$ and $\Omega_0\equiv L^{-2}\alpha$ are integration constants appearing in \eqref{metric}, $L$ is the AdS radius, and $\kappa$ is the reduced Newton constant. Thus, \eqref{metric} is the exact holographic dual to this setting. Non-spinning models of plasma balls can be found in \cite{Nastase:2005rp, Emparan:2009dj, Aharony:2005bm}.

Plasma balls have a natural maximum size, which is determined by the region where their boundary reaches the speed of light, namely at $\gamma=\infty$. This physical boundary of the system extends into the bulk as a new conformal boundary, located in the bulk at $y=\infty$. When $y$ is restricted to the Minkowski boundary \eqref{b1} $y=\gamma$. Hence, the bulk metric \eqref{metric} has two energy scales: the holographic renormalization-group flow, parameterized by $r$, and the Lorentzian energy scale, parameterized by $y$. At the conformal boundary located at $y=\infty$, we find the following metric:
\begin{equation} \label{b2}
g^y_{\mu \nu}d\hat{z}^{\mu}d\hat{z}^{\nu}=-f(r)dt^2+\frac{dr^2 L^2}{r^2 \alpha^2 F(r)}+\frac{L^4}{\alpha^2}\left(2+\frac{1}{r^2}-f(r)\right)d\phi^2-2\frac{(mr-q^2)L^4}{\alpha r^2}dtd\phi\;,
\end{equation}
where $F(r)=L^{-2}r^4-mr+q^2+L^{-2}r^2$ and $f(r)=1-\frac{mL^2}{r}+\frac{q^2L^2}{r^2}$. The VEV of the holographic energy--momentum tensor at this boundary vanishes. The boundary dual graviton coupled to the spinning disk \eqref{EMtensor} is described by a metric $*g^r$ satisfying \cite{deHaro:2008gp, Bakas:2008gz, Mukhopadhyay:2013gja}
\begin{equation}
\frac{L^2}{\kappa} C_{\mu \nu}(*g^r)=\pm\langle T_{\mu \nu} \rangle,
\end{equation}
where $C_{\mu \nu}$ is the Cotton tensor.\footnote{We use the convention
\[
C_{\mu \nu}
=
\epsilon_{\mu}{}^{\eta \xi}
\nabla_{\eta}
\left(
R_{\nu \xi}
-\frac{1}{4}g_{\nu \xi}R
\right),
\]
where $\epsilon_{\mu\nu\rho}$ is the Levi--Civita tensor.}
One can verify that the solution corresponding to the plus sign\footnote{The minus sign is obtained by taking $r\rightarrow -r$ in \eqref{b2}.} is
\begin{equation}
*g^r=g^y(
t\rightarrow \mathrm{i}t,
\phi\rightarrow \mathrm{i}\phi,r \rightarrow \gamma) \, .
\end{equation}
Namely, one needs to do a double Wick rotation and identify the holographic renormalization-group scale with the Lorentz factor,
\begin{equation}
r\leftrightarrow\gamma.
\end{equation}
We therefore conclude that fixing the conformal boundary at $y=\infty$ is equivalent to fixing the double Wick-rotated dual graviton associated with the Minkowski boundary. Incidentally, this solves the open problem of constructing the dual graviton of a spinning disk. The electromagnetic duality between the two boundaries is perfectly realized in the gauge field, as the VEVs of the currents associated with the bulk $U(1)$ Maxwell field are exchanged with those of their dual currents when computed at the boundaries \eqref{b1} and \eqref{b2}.

The double Wick rotation is required because the vector $\zeta=\partial_t+\Omega_0\partial_{\phi}$, which is timelike in the interior of the plasma disk, becomes null when the plasma reaches the speed of light and spacelike beyond that point. Indeed,
\begin{equation}
g^y_{\mu \nu}\zeta^{\mu}\zeta^{\nu}>1 \, ,
\end{equation}
so the different conformal boundaries are naturally classified by the norm of $\zeta$, where $t$ and $\phi$ are common coordinates to all of them. Moreover, there is a codimension-two conformal boundary with representative metric
\begin{equation}
-dt^2+\Omega_0^{-2}d\phi^2
\end{equation}
that joins $g^r$ and $g^y$. The different conformal boundaries are thus reminiscent of the causal structure of a black hole, with the exterior region represented by $g^y$. The metric \eqref{b2} contains an integration constant $q$ that is absent from the original conformal background, though it appears in the sourceless
$U(1)$ gauge field as seen in eq. (\ref{A}).
When $m=0$ and $q\neq 0$, \eqref{b2} is locally $AdS_3$ with AdS radius $\frac{L}{q\alpha}$. The constant $q$ is the duality-invariant part of the $U(1)$ gauge field \eqref{A}. Hence, the dual graviton of the vacuum, namely the sector in which the energy--momentum tensor vanishes, of a CFT coupled to a global $U(1)$ current in Minkowski spacetime is $AdS_3$ with an AdS radius proportional to the inverse of the charge. This prediction cannot be extracted from the Cotton tensor prescription, since the Cotton tensor vanishes on both boundaries in this case. It is remarkable that the vacuum structure, namely the state satisfying $\langle T_{\mu\nu}\rangle=0$, of a strongly coupled QFT in Minkowski spacetime coupled to a global $U(1)$ current is nevertheless nontrivial, provided the state is rotating. This observation may open a fruitful direction to study in quantum field theories.

Note that in this $m=0, q\neq 0$ case, in the bulk 
$T_{\mu\nu}\neq 0$. Reversely, if $q=0$ but $m\neq 0$, $\langle T_{\mu\nu}\rangle\neq0$
in the CFT, and one still has rotation. This is consistent with the holographic definition
of the Mach's principle in general relativity proposed in \cite{Khoury:2006hg}, in which motion and non-inertial frames, 
so in particular rotation, are 
only defined with respect to matter, either in the bulk or on 
a boundary (that acts as a kind of gravitational Faraday cage, 
trading boundary conditions for boundary matter).

The bulk solution also has several remarkable properties that are interesting beyond the holographic context. First, it is completely free of curvature singularities. Furthermore, there is a region of parameter space in which there are no closed timelike curves (CTCs), either outside or inside the horizons. Super-extremal configurations are therefore smooth solitons. This motivates us to investigate whether these solitons correspond to stable ground states. We answer this affirmatively by showing that both the purely ``magnetic'' and purely ``electric'' families of solutions admit supersymmetric configurations.

The remainder of this paper is   organized as follows. In Sec.~2 we present the Einstein--Maxwell theory together with the supersymmetry transformations relevant for our analysis. Section~3 introduces the new family of exact solutions and studies their global properties, including regularity, horizons, and the conditions for the absence of closed timelike curves. In Sec.~4 we discuss the supersymmetric limits, showing that both the electric and magnetic branches admit smooth BPS solitons. Section~5 shows how the static planar Reissner--Nordström--AdS black hole is recovered as an appropriate limit of the rotating solution. In Sec.~6 we analyze the conformal boundaries of the spacetime. Section~7 develops the holographic interpretation, including the spinning plasma on the Minkowski boundary and the dual-graviton description associated with the second boundary. Electromagnetic duality and its realization as an exchange between the two boundary theories are presented in Sec.~8. The last section has conclusions.

\section{Setup}

The action is
\begin{equation}
S[g,A]=\frac{1}{2\kappa}\int d^4x \sqrt{-g}\left[R+\frac{6}{L^2}-\frac{1}{4} F^2\right]+S_{\partial}\;,
\end{equation}
where $F(A)_{\mu\nu}=\partial_{\mu}A_{\nu}-\partial_{\nu}A_{\mu}$. $S_{\partial}$ contains boundary terms so the action is well defined. The field
equations are%
\begin{align}
&  \partial_{\mu}\big(\sqrt{-g}F^{\mu\nu}\big)=0\,\,,\nonumber\\[3pt]
&  \nonumber R_{\mu\nu}-\frac{1}{2}g_{\mu\nu}R-\frac{1}{2}\big[F_{\mu\rho}\,F_{\nu}%
{}^{\rho}-\frac{1}{4}g_{\mu\nu}F_{\rho\sigma}F^{\rho\sigma}\big] -\frac
{3}{L^{2}}\,g_{\mu\nu}=0\,.\\
\end{align}
When supplemented by the fermionic sector, the theory is invariant
under supersymmetric transformations of all the fields\cite{Freedman:1976aw, Fradkin:1976xz,deWit:1980lyi,deWit:1984wbb}. We only make use of the integrability condition of transformation of the Rarita-Schwinger fields
$\psi_{\mu}{\!}^{i}\,$:

\begin{equation}\label{eq:gravitinovariation}
\delta\psi_{\mu}{\!}^{i}=\;  2\,\big(\partial_{\mu}+\tfrac{1}{4}%
\omega_{\mu}{\!}^{ab}\gamma_{ab}\big)\epsilon^{i}-\tfrac{1}{L}A_{\mu
}\,t^{i}{\!}_{j}\,\epsilon^{j}-\tfrac{1}%
{4}F(A)_{\rho\sigma}\gamma^{\rho\sigma}\gamma_{\mu}\,\varepsilon
^{ij}\,\epsilon_{j}+L^{-1}\,\varepsilon^{ij}\,t_{j}{}^{k}\,\gamma_{\mu
}\epsilon_{k}\,,
\end{equation}
where $i=1,2$. 
\section{The Geometry}
We constructed the solution inspired by the Carter class of solutions \cite{Carter:1968ks} and by the requirement that the conformal boundary be of Minkowski type \eqref{b1}, with the boundary matter described by a spinning perfect fluid \eqref{EMtensor},

\begin{align}\nonumber
ds^2& =
\alpha^2
\left(
\frac{
y^4 r \left(m-\frac{q^2}{r}\right)
}{
y^2+r^2
}
-\frac{y^2 r^2}{L^2}
\right)dt^2
-
\frac{
2 \alpha y^2 r (y^2-1)
\left(m-\frac{q^2}{r}\right)L^2
}{
y^2+r^2
}
\,dt\,d\phi\\\nonumber
&+
\left(
\frac{
r (y^2-1)^2
\left(m-\frac{q^2}{r}\right)L^4
}{
y^2+r^2
}
+
L^2 (y^2-1)(r^2+1)
\right)d\phi^2\\ \label{metric}
&+
\frac{
L^2(y^2+r^2)
}{
y^2 (y^2-1)
}
dy^2
+
\frac{
y^2+r^2
}{\frac{r^4}{L^2}+\frac{r^2}{L^2}
-m r+q^2
}
dr^2
\end{align}
with the gauge field 

\begin{equation}\label{A}
A
=
\left(
\frac{
2q r y
\left(
y\cos\theta
+
r\sin\theta
\right)
}{
r^2+y^2
}
\,\alpha
-C_1
\right)dt
-
\left(
\frac{
2q L^2
\left(
-r\cos\theta
+
y\sin\theta
+
r y^2 \cos\theta
+
y r^2 \sin\theta
\right)
}{
r^2+y^2
}
-C_2\right)d\phi
\end{equation}
where $C_1$ and $C_2$ are constants that ensure the gauge field is well defined at the degeneration loci of the Killing vectors. $\theta$ is an integration constant that is related to electromagnetic duality and that is present only in the gauge field and not in the metric. The coordinate system is such that the conformal boundary at $r\rightarrow \infty$ is static.

The range of the coordinates is 
\begin{equation}\label{range}
t\in \mathbb{R}\, ,\quad \phi \in (0,2\pi)\, ,\quad y \in (1,\infty),\quad r \in \mathbb{R} \, .
\end{equation}
The parameterization of the solution is chosen so that the degeneration surface of $\partial_{\phi}$, at $y = 1$, yields a canonically normalized $\phi$. The determinant of the metric is
\begin{equation}
\det g = -\alpha^2 L^4 (r^2+y^2)^2\, .
\end{equation}
Therefore, given the coordinate range \eqref{range}, $\det g$ never vanishes, and the metric is invertible everywhere. Furthermore, a simple inspection shows that the gauge field diverges only at $r^2 + y^2 = 0$ and all the metric coefficients are analytic everywhere, except when there is a degeneration surface of a Killing vector. Consequently, the spacetime is free of curvature singularities. 

The function
\begin{equation}
F(r)=g_{rr}^{-1}(r^2+y^2)=L^{-2}r^4-mr+q^2+L^{-2}r^2\, ,
\end{equation}
defines the location of the event horizon. 
There are two event horizons at $F(r)=0$. Let us call them $r_0$ and $r_1$. It is convenient to express $m$ and $q$ in terms of these roots,
\begin{equation}\label{m}
m
=
\frac{(r_1+r_0)\left(r_0^2+r_1^2+1\right)}{L^2},
\qquad
q^2
=
\frac{
r_0 r_1
\left(
r_1^2+r_0 r_1+1+r_0^2
\right)
}{
L^2
}.
\end{equation}
The event horizons have generators of the form
\begin{equation}
\xi_i= \partial_t+\omega_i\partial_{\phi}\, , \qquad \omega_i = \frac{\alpha r_i^2}{L^2(1+r_i^2)}\, , \quad i=0,1 \, .
\end{equation}
We take $r_0>r_1$. When there is no charge the $r_1=0$, which in the massless limit is the location of the Poincaré horizon. The horizon located at $r=r_0$ has temperature and area density
\begin{align}
T=\frac{\alpha\left(3r_0^4+r_0^2-q^2L^2\right)}{4\pi r_0(r_0^2+1)L^2}\, ,\\
\mathcal{A}=\sqrt{g_{yy}g_{\phi \phi}}=L^2(r_0^2+1).
\end{align}
Regularity of the gauge field at $y=1$ and $r=r_0$ fixes the constants in \eqref{A} as
\begin{equation}
C_1
=
\frac{
2 r_0 \alpha q
\left(
r_0\sin\theta+\cos\theta
\right)
}{
r_0^2+1
},
\qquad
C_2
=
2qL^2\sin\theta \, .
\end{equation}
Since $r\in\mathbb{R}$, the topology of the solution is that of a cylinder, with the conformal boundary \eqref{b2} forming its lateral surface at $y\to\infty$, and the flat boundaries \eqref{b1} capping the cylinder at $r\to\pm\infty$. These two asymptotic Minkowski boundaries rotate in opposite directions. Indeed, the sense of rotation of the spacetime reverses at the radius where $g_{t\phi}=0$, namely
\[
r_{*}=\frac{q^2}{m}\  .
\]
For $r>r_{*}$, the spacetime rotates in one direction, with the rotation vanishing as $r\to\infty$. For $r<r_{*}$, the sense of rotation is reversed, and its magnitude again decreases to zero as $r\to-\infty$. In the solitonic case, namely when $F(r)$ is everywhere positive, the solution resembles a three-dimensional rotational vortex centered on the smooth origin of the disk, located at $y=1$ and $r=0$.

\subsection{No closed timelike curves}
There are no closed timelike curves provided $g_{\phi\phi}$ is non-negative,
\begin{align}\label{gpp}
g_{\phi\phi}
&=
L^4\frac{y^2-1}{y^2+r^2}\big(h(r)y^2+F(r)\big)\geq 0\, ,
\\ \label{gp2}
h(r)
&=
(r^2+1)^2L^{-2}-F(r)\, .
\end{align}
Equation \eqref{m} allows us to write $h(r)$ in a manifestly positive form,
\begin{equation}
(r^2+1)^2L^{-2}-F(r)
=
\frac{(r-r_0)^2}{L^2}
+\frac{r_0^4+q^2L^2+3r_0^2}{L^2r_0}(r-r_0)
+\frac{(r_0^2+1)^2}{L^2}\, .
\end{equation}
This expression is positive for $r>r_0$, and therefore there are no CTCs in this region. In the region $r_0>r>r_1$, one has $F(r)=-|F(r)|$ and $h(r)>0$, with a possible zero of $g_{\phi\phi}$ at $y=y_0$, where
\begin{equation}
y_0^2
=
-\frac{F(r)}{h(r)}
=
\frac{|F(r)|}{|F(r)|+(r^2+1)^2L^{-2}}
<1\, ,
\end{equation}
which lies outside the allowed coordinate range. It follows that this region is also free of CTCs.

To exclude closed timelike curves for $r<r_1$, it is necessary to require that
\begin{equation}\label{quadratic}
h(r)
=
mr+L^{-2}r^2-q^2+L^{-2}
>0\, ,
\end{equation}
otherwise $g_{\phi\phi}$ can become negative for sufficiently large values of $y$. This condition is guaranteed if the discriminant of the quadratic polynomial \eqref{quadratic} is negative, which yields
\begin{equation}\label{CTC}
m^2L^4+4q^2L^2-4<0\, .
\end{equation}

Furthermore, it is necessary to verify that the condition \eqref{CTC} is compatible with the existence of horizons. We have checked this numerically, as shown in Fig.~1. The result is that there exist open regions of parameter space containing either black holes or solitons that are free of CTCs. 

The solitons arise whenever $F(r)>0$ for all $r$. Since horizons are absent in this case, the absence of closed timelike curves is ensured by \eqref{CTC}. Hence, the region of parameter space satisfying
\begin{equation}
F(r)>0 \qquad \text{and} \qquad m^2L^4+4q^2L^2-4<0
\end{equation}
corresponds to smooth horizonless geometries that are free of both curvature singularities and causal pathologies.

\begin{figure}[t]
\centering
\includegraphics[width=0.7\textwidth]{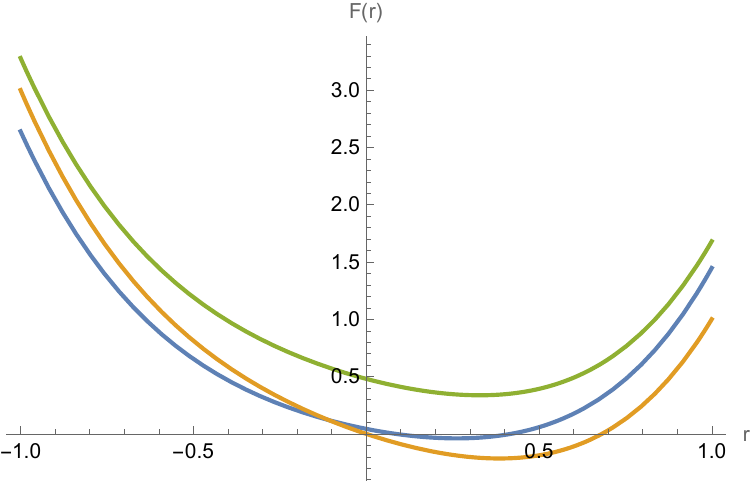}
\caption{Parameter-space regions corresponding to black holes and solitons free of closed timelike curves. The green line is a soliton and the other two are black holes. From top to bottom the parameters are $(m=0.8, q^2=0.48)$, $(m=0.596, q^2=0.0435)$, $(m=1, q^2=0)$, measured in units of $L^{-2}$. }
\label{fig:CTC}
\end{figure}

\section{Supersymmetry}
We find that the integrability of the supersymmetry variations has a vanishing determinant in the following cases
\begin{align}
\theta = &0 \implies mL=\pm 2q\, , \\ 
\theta = &\frac{\pi}{2} \implies m=0\, ,\qquad qL=\frac{1}{2} .
\end{align}
In both of these cases $F(r)>0$ everywhere. When $\theta = \frac{\pi}{2}$ we find that $h(r)>0$ so there are no CTC. When $\theta = 0$ we find that is necessary that $2q^2L^2<1$ to have no CTC. For the technical details of the supersymmetry analysis see \cite{Caldarelli:1998hg} and references therein.

\section{The static limit}

The static black hole solution is recovered with the following change of coordinates and reparameterization

\begin{equation}
r=\frac{\rho}{\alpha(1+\frac{1}{2L^2}\alpha^2 \hat{x}^2)}\, ,\quad  y=1+\frac{1}{2L^2}\alpha^2 \hat{x}^2\, ,\quad m=\frac{m_s}{\alpha^3}\, ,\quad q=\frac{q_s}{\alpha^2}\, ,\quad r_0=\frac{\rho_0}{\alpha}\, ,
\end{equation}
which yields the standard static dyonic planar black hole when $\alpha\rightarrow 0$,
\begin{align}
ds_s^2&=-f(\rho)dt^2+\frac{d\rho^2}{f(\rho)}+\rho^2(d\hat{x}^2+\hat{x}^2d\phi^2)\, ,\qquad f(\rho)=\frac{\rho^2}{L^2}-\frac{m_s}{\rho}+\frac{q_s^2}{\rho^2}\, ,\\
A&=
\frac{2q_s\cos\theta}{\rho\rho_0}
(\rho_0-\rho)\, dt
-
q_s\sin\theta\, \hat{x}^2\, d\phi \, .
\end{align}
The spinning solution yields a locally constant curvature geometry when $m=0=q$ (for general $\alpha$). To show this, we replace the change of coordinates
\begin{equation}
\hat{\rho}=r y \alpha\, , \qquad \hat{w}=L \frac{\sqrt{(y^2-1)(r^2+1)}}{\alpha\, y\, r}\, , 
\end{equation}
in the metric \eqref{metric} with $m=0=q$. It yields a pure $AdS_4$ metric,
\begin{equation}
ds^2=-\frac{\hat{\rho}^2}{L^2}dt^2+L^2 \frac{d\hat{\rho}^2}{\hat{\rho}^2}+\hat{\rho}^2(d\hat{w}^2+\hat{w}^2d\phi^2) 
\end{equation}

\section{The conformal boundary}

The conformal boundary of the spinning solution is non-trivial. For $r>0$ we find two conformal boundaries, with $\omega=\frac{L}{\alpha r y}$. The usual Minkowski boundary at the UV of the holographic coordinate \eqref{b1},
\begin{equation}\label{rbound}
ds_r^2=\lim_{r\rightarrow \infty} \omega^2 ds^2=-dt^2+dw^2+w^2d\phi^2 = g^r_{\mu \nu}dz^{\mu}dz^{\nu}\, , \qquad y=\frac{1}{\sqrt{1-\alpha^2L^{-4}w^2}}\;,
\end{equation}
and a spinning black hole or spinning soliton
at the conformal boundary $y\to \infty$ \eqref{b2},
\begin{align} \label{otherboundary}
ds_y^2&=\lim_{y\rightarrow \infty} \omega^2 ds^2=g^y_{\mu \nu}d\hat{z}^{\mu}d\hat{z}^{\nu}\;.
\end{align}
We also find that a codimension-two conformal boundary can be obtained from the bulk by setting $r=y$ and taking the corresponding asymptotic limit,
\begin{equation}
ds_{ry}^2=\lim_{y\rightarrow\infty}\omega^2(r=y)\,ds^2
=-dt^2+\Omega_0^{-2}\,d\phi^2\,,
\end{equation}
which coincides with the Minkowski boundary at the locus where the fluid rotates at the speed of light.
\subsection{The boundary at $y=\infty$}

This geometry deserves some comments. When it is a black hole we checked it has the same temperature and angular velocity than the bulk black hole. When $m=0$ its Riemann tensor yields
\begin{equation}
R^{\mu \nu}_{\xi \eta}=-\frac{q^2 \alpha^2}{L^2}(\delta^{\mu}_{\xi}\delta^{\nu}_{\eta}-\delta^{\mu}_{\eta}\delta^{\nu}_{\xi}) \, ,
\end{equation}
Thus, the geometry is locally that of three-dimensional AdS spacetime with curvature radius $L/(q\alpha)$. For the supersymmetric solution with $m=0$, the corresponding AdS$_3$ radius is

\begin{equation}\label{AdS3}
\frac{L}{q\alpha}=\frac{2}{\Omega_0}\, .
\end{equation}

This relation suggests the existence of a supersymmetry-protected duality. In particular, the vacuum sector of a strongly coupled quantum field theory, namely, the sector with vanishing expectation value of the energy-momentum tensor, at finite angular velocity $\Omega_0$ and in the presence of a purely magnetic field is dual to a quantum field theory in AdS$_3$, with AdS radius given by Eq.~\eqref{AdS3}, the same angular velocity, and a purely electric field of the same magnitude as the magnetic field. This seems to be a refinement of the holographic description of mirror symmetry to the case in which there is finite angular velocity, see \cite{Witten:2003ya}.

When $m\neq 0$ we get that the Cotton tensor is not zero,
\begin{equation}
\frac{L^2}{\kappa}C_{\mu \nu}=-\frac{3}{2}\hat{\rho}V_{\mu}V_{\nu}+\frac{\hat{\rho}}{2}g^y_{\mu \nu}\, , \qquad V^{\mu}=r(\partial_t+\Omega_0\partial_\phi)\, , \qquad \hat{\rho}= \frac{L^4}{\kappa}r^3 m \Omega_0^3  
\end{equation}
where $V_{\mu}V^{\mu}=1$ and we use $\alpha\equiv L^2\Omega_0$. This means that the geometry is singular at $r=\infty$ which is due to the conformal frame and not a property of the four dimensional geometry. We shall see that the singularity is natural due to the holographic interpretation.

\section{Holographic analysis}

We proceed to implement the standard holographic renormalization of \cite{Balasubramanian:1999re} at the Minkowski boundary, the VEV of the energy momentum tensor  computed in terms of the metric $h_{\mu \nu}=g_{\mu \nu}-N_{\mu}N_{\nu}$, with $N_{\mu}=\delta_{\mu}^{r}\sqrt{g_{r r}}$ and the extrinsic curvature $K_{\mu \nu}=\frac{1}{2}(\nabla_{\mu}N_{\nu}+\nabla_{\nu}N_{\mu})$, its trace $K$ and the Einstein tensor of $h$,  giving
\begin{equation}
\langle T_{\mu \nu} \rangle=\lim_{r\rightarrow \infty} \frac{1}{\omega \kappa}(L G_{\mu \nu}(h)-\frac{2}{L}h_{\mu \nu}-K_{\mu \nu}+h_{\mu \nu}K)\, .
\end{equation}
The calculation yields a perfect fluid,
\begin{equation}
\langle T^r_{\mu \nu} \rangle=U_{\mu}U_{\nu}(\rho+P)+g^{r}_{\mu \nu}P\;,
\end{equation}
with 
\begin{equation}
P=\frac{\rho}{2}=\frac{L^4}{2\kappa}\gamma^3 m \Omega_0^3\, , \qquad U=\gamma(\partial_{t}+\Omega_0\partial_{\phi})\, , \qquad U^{\mu}U_{\mu}=-1\, , \qquad \gamma=\frac{1}{\sqrt{1-\Omega_0^2 w^2}}\, . 
\end{equation}
Hence, there is a disk of strongly coupled conformal plasma spinning with angular velocity $\Omega_0$ at this boundary of the spacetime. We note that the coordinate range $y\in (1,\infty)$ maps to $w\in (0,\Omega_0^{-1})$. Namely, the geometry ends naturally when the boundary of the fluid
disk is spinning at the speed of light. This geometry has 
vanishing Cotton tensor.

We should pause here to observe that this
definition of the
angular velocity of the spinning fluid, $\Omega_0$,
is not what we usually have as the thermodynamical angular velocity of the CFT. Usually, for example 
for the case of the Kerr-Newman-AdS solutions previously 
studied in holography, we define, for an inertial observer, the angular velocity of the CFT, $\Omega_{CFT}$, as the 
difference between the angular velocity of the horizon 
and that of the boundary coordinate system \cite{Hawking:1998kw,Hawking:1999dp,Papadimitriou:2005ii}. In our case, the angular velocity of the $r\rightarrow \infty$ boundary coordinate system is zero, $\Omega_{\infty}=0$, since we obtained a Minkowski conformal boundary in a static coordinate system, so that definition gives $\Omega_{CFT}=\omega
_{\rm horizon}-\Omega_{\infty}=\omega_0\neq \Omega_0$. This difference between the fluid angular velocity and that of the microscopic description has been known for a while \cite{Bhattacharyya:2007vs}.

The same energy-momentum tensor calculation 
in the other boundary \eqref{otherboundary} yields 
\begin{equation}
\langle T^y_{\mu \nu} \rangle =0 \, .
\end{equation}
However, the dual energy momentum tensor \cite{deHaro:2008gp, Bakas:2008gz, Mukhopadhyay:2013gja}
yields 
\begin{equation}
\langle *T^y_{\mu \nu} \rangle \equiv\frac{L^2}{ \kappa}C_{\mu \nu}=-\frac{3}{2}\hat{\rho}V_{\mu}V_{\nu}+\frac{\hat{\rho}}{2}g^y_{\mu \nu} \, .
\end{equation}
We find that the energy density of the dual energy-momentum tensor is the same as on the Minkowski boundary, provided we identify the holographic energy scale with the Lorentz factor, $r=\gamma$, and measure it with respect to the timelike vector $\hat{U}=\mathrm{i}V$, namely
\begin{equation}
\langle *T^y_{\mu \nu} \rangle\hat{U}^{\mu}\hat{U}^{\nu}_{r=\gamma}=\langle T^r_{\mu \nu} \rangle U^{\mu}U^{\nu}\, .
\end{equation}

Indeed, the Minkowski boundary describes a spinning fluid whose elements follow trajectories tangent to \(U\). As the fluid velocity approaches the speed of light, these trajectories become null. Consequently, the region at \(y=\infty\) but finite \(r\) lies beyond this light surface, where the Killing vector \(\partial_t+\Omega_0\partial_\phi\) is spacelike. We therefore normalize it to unity through the definition of \(V\). The singularity of the geometry at \(r=\infty\) in this conformal frame reflects the singular behavior of the energy-momentum tensor at the light surface, where the fluid reaches the speed of light. 

The geometry defined by the boundary graviton $g^y_{\mu \nu}$ has a Cotton tensor that is equal to the energy momentum tensor of the Minkowski boundary provided that the complex diffeomorphism is applied
\begin{equation}
t\rightarrow \mathrm{i} t\, , \qquad \phi \rightarrow \mathrm{i} \phi \, , \qquad r \rightarrow y \, .
\end{equation}
Hence, fixing $g^y_{\mu \nu}$ is equivalent to fixing the dual graviton of $g^r_{\mu \nu}$. The standard energy momentum tensor vanishes at $y=\infty$ and therefore that boundary satisfies either the Dirichlet boundary condition or the Neumann one. 

\section{Electromagnetic duality}

As we have seen, electromagnetic duality is associated with the interchange of the coordinates $r$ and $y$. Accordingly, in this section we present the results in Minkowski spacetime expressed in terms of the coordinate $y$, \eqref{rbound},
\begin{equation}
ds_r^2=-dt^2+\Omega_0^{-2}\frac{dy^2}{(y^2-1)y^4}+\Omega_0^{-2}(y^2-1)\frac{d\phi^2}{y^2}\, .
\end{equation}
The determinants of the metrics in the different boundaries are 
\begin{equation}
\det g^r=-\Omega_0^{-4}y^{-6}\, , \qquad \det g^y=-\Omega_0^{-4}r^{-6}\
\end{equation}

In the standard quantization the boundary value of the gravity fields yields a VEV to a current in the QFT. The VEVs of the conserved currents are 
\begin{equation}
\langle J_i^{\nu} \rangle = -\frac{1}{\sqrt{-\det g^i}}\lim_{i\rightarrow \infty}\frac{L}{2\kappa} \sqrt{-h} N^i_{\mu} F^{\mu \nu}\, , \qquad \langle \tilde{J}_i^{\nu} \rangle = -\frac{1}{\sqrt{-\det g^i}}\lim_{i\rightarrow \infty}\frac{L}{2\kappa} \sqrt{-h} N^i_{\mu} \tilde{F}^{\mu \nu}\, .
\end{equation}
where $i=r,y$ depending on the boundary, $N^i=\sqrt{g_{ii}}$. We define the dual currents, $\tilde{J}$, with $\tilde{F}_{\mu \nu}=\frac{1}{2}\epsilon_{\mu \nu \lambda \sigma}F^{\lambda \sigma}$.
We find that the currents are proportional to the fluid velocity field $U$,
\begin{equation}
\langle J_r \rangle=-\frac{L^3 q \Omega_0^2 y^3 \cos\theta}{\kappa}\left(\partial_t+\Omega_0\partial_{\phi} \right) \qquad \langle \tilde{J}_r \rangle=\frac{L^3 q \Omega_0^2 y^3 \sin\theta}{\kappa}\left(\partial_t+\Omega_0\partial_{\phi} \right)
\end{equation}
and
\[
\qquad
\begin{pmatrix}
0&1\\
-1&0
\end{pmatrix}\begin{pmatrix}
J^y\\
\tilde{J}^y
\end{pmatrix}=\begin{pmatrix}
J^r(y=r)\\
\tilde{J}^r(y=r)
\end{pmatrix}.
\]
Hence, we get that the currents and their EM duals at the boundaries are related by bulk EM duality plus the exchange
\begin{equation}
r\leftrightarrow y\, .
\end{equation}
The gauge field at the boundaries are 
\begin{align}\label{Asless}
A^r&\equiv\lim_{r\to\infty}A=\left(2q\a \sin\theta\; (y-1)
+\frac{2q\alpha\left(\sin\theta - r_0\cos\theta\right)}{r_0^2+1}
\right)dt-2q L^2 \sin\theta\; (y-1)d\phi\, ,\\
C^r&\equiv
\left(\frac{2q\alpha\left(\sin\theta - r_0\cos\theta\right)}{r_0^2+1}-2q\alpha \sin \theta\right)  dt+2q L^2 \sin\theta d\phi\, ,\\
A^y&\equiv\lim_{y\to\infty}A=\left(2q\a\cos\theta \; r
-\frac{2\alpha q r_0\left(r_0\sin\theta+\cos\theta\right)}{r_0^2+1}
\right)dt+2q L^2(\sin\theta-\cos\theta\; r) d\phi\, , \\
C^y&\equiv
-\frac{2\alpha q r_0\left(r_0\sin\theta+\cos\theta\right)}{r_0^2+1}
dt+2q L^2\sin\theta d\phi\, , 
\end{align}
where we have also separated the constant part of the gauge fields in $C^i$. Both satisfy the Lorenz gauge $\d^\mu A^i_\mu=0$, 
and a kind of Coulomb gauge, $A^r_r=0=A^y_y$ and its non-constant part is orthogonal to $U$, $A^i_{\mu}U^{\mu}=C^i_{\mu}U^{\mu}$. $F^i=dA^i$ is also orthogonal to $U$ and hence it is purely magnetic {\em in the comoving frame}
in the Minkowski region ($r\to \infty$) and purely electric 
(also in the ``comoving frame'') in the black hole region ($y\to \infty$), and we find that the gauge fields seem to have a position dependent mass and topological mass at the boundary. They satisfy the following equation\footnote{We use the convention that the Levi-Civita tensor is proportional to $\sqrt{-\det g}$.}
\begin{equation}
\nabla_{\nu}F^{ i\nu \mu}=-M^2_i \epsilon^{\mu \lambda \eta}F^i_{\lambda \eta}+m_i^2 (A^{i \mu}-C^{i \mu})\, ,
\end{equation}
\begin{equation}
m_r^2=\Omega_0^2 y^2(y^2-1)\, ,\quad M^2_r=y^2\Omega_0\, ,\quad m_y^2=\Omega_0^2 L^2 F(r)\, ,\quad M^2_y=r^2 \Omega_0\, .
\end{equation}
where $\epsilon^{\mu \lambda \eta}$ is the Levi-Civita tensor of the corresponding boundary. It is possible to see that the position dependent masses blow up at the common
boundary (with the disk).
The resulting $F=dA$ gives the electric 
and magnetic fields for the plasma disk and black hole boundaries, are naturally given in terms of the three-dimensional fields. We find that
\begin{align}
\frac{L}{4\kappa}\epsilon^{\mu \lambda \eta}F^r_{\lambda \eta} &=\frac{L^3}{\kappa}y^3q\sin \theta \Omega_0^2(\partial_t+\Omega_0\partial_{\phi})=\tilde{J}^r\, ,\\
\frac{L}{4\kappa}\epsilon^{\mu \lambda \eta}F^y_{\lambda \eta} &=\frac{L^3}{\kappa}r^3q\cos \theta \Omega_0^2(\partial_t+\Omega_0\partial_{\phi})=-\tilde{J}^y\,,
\end{align}
so the dual currents $\tilde J$ can be derived from a CS term
via $\delta S_{\rm CS}/\delta A_\mu$, 
just like the Integer Quantum Hall Effect (IQHE) currents. 
However, using
the physical gauge fields, $A_{\rm phys}=L^{-1}A$, and $L,\kappa$
from the $AdS_4\times S^7$ M-theory solution, $R_7=l_P(32 \pi^2N)
^{1/6}, R_4=R_7/2$, and $2\kappa=(2\pi)^8l_P^93/[\pi^4(R_7)^7]$, 
we obtain the (dual) Hall conductivity
\be
\tilde \sigma_H=\frac{L^2}{4\kappa}=\frac{\sqrt{2}}{16\pi 
}N^{3/2}\;,
\ee
unlike the $N/(2\pi)$ of IQHE, arising from the CS term 
quantization.
Note that, from the point of view of the boundary CFTs, the
electric and magnetic fields are just external fields, as 
usual in holography. Nevertheless, we can obtain a certain notion of 
electric-magnetic duality acting on these external fields, by 
considering $F$ and $\tilde{F}$ at each boundary. 
\[
\qquad
\begin{pmatrix}
0&1\\
-1&0
\end{pmatrix}\begin{pmatrix}
F^y\\
\tilde{F}^y
\end{pmatrix}=\begin{pmatrix}
F^r(y=r)\\
\tilde{F}^r(y=r)
\end{pmatrix},
\]
Hence, we get the gauge fields and their duals at the boundaries are related by EM duality plus the exchange
\begin{equation}
r\leftrightarrow y\, .
\end{equation}

Thus, these two theories are related by Mirror symmetry along the lines described by \cite{Witten:2003ya}. The purely magnetic theory is that at $r=\infty$ and the purely electric theory is that at $y=\infty$.

\section{Conclusions}

In this paper, we constructed a new class of solutions to the four-dimensional Einstein--Maxwell equations with a cosmological constant. These solutions provide a holographic description of strongly coupled three-dimensional spinning matter. The solutions remarkable have no curvature singularity and no CTCs in a region of the parameter space. 

Rigidly spinning matter has a maximum spatial extent, determined by the locus at which the tangential velocity reaches the speed of light. We found that this natural boundary extends into the bulk as a second conformal boundary. Our results demonstrate that holography can be consistently formulated in this multi-boundary setting. Furthermore, we provide a precise interpretation of the metric induced on this second boundary, showing that it is obtained from the dual graviton by a double Wick rotation. The resulting boundary-boundary duality also maps the expectation values of conserved currents to the boundary values of the corresponding gauge fields. We believe that one of the central outcomes of this construction is the realization that the Lorentz scale is the electromagnetic dual of the renormalization-group flow.

To describe a plasma disk that terminates at some radius $R<\Omega_0^{-1}$, the spacetime must be truncated at the corresponding value of $y$, so that the bulk geometry acquires a boundary rather than a conformal boundary in the $y$ direction. A holographic description of CFTs with boundaries has been proposed in \cite{Takayanagi:2011zk}. At the cutoff surface, one imposes Neumann boundary conditions, which relate the induced energy--momentum tensor to that of an end-of-the-world brane. This framework appears to fit naturally with our construction, since the energy--momentum tensor vanishes at $y=\infty$, corresponding to the limit in which the end-of-the-world brane disappears. One may therefore regard the boundary at $y=\infty$ as satisfying a Neumann boundary condition, while the standard Dirichlet boundary condition is fixed at $r=\infty$. 

We found a large class of horizonless solitonic solutions, providing supergravity states that may account for the microscopic origin of the entropy of the corresponding spinning black holes. These spinning solitons exist above stable supersymmetric ground states and, heuristically, appear to be connected to the fluid described by the Nielsen--Olesen vortex \cite{Nielsen:1973cs}, which has long been proposed as a candidate for the vacuum of strongly coupled gauge theories. 

The everywhere regular supersymmetric solutions are the ground states of the system at fixed charges. This result might resolve a longstanding problem concerning the stability of black branes in supergravity. As first noted by Romans \cite{Romans:1991nq}, the supersymmetric limit of static planar black holes is horizonless, yielding naked singularities. Since this theory is a consistent truncation of M-theory compactified on $S^7$, the same was later shown to hold for purely electric black holes in the dilatonic $U(1)^4$ sector, whereas magnetically charged black holes were found to admit regular supersymmetric limits \cite{Cacciatori:2009iz}. Independently, it was pointed out that electrically charged black branes are unstable, raising the question of the endpoint of the instability, particularly given that the supersymmetric limit is singular \cite{Cvetic:1999rb,Gubser:2000ec}. Indeed, even the planar Reissner--Nordström black hole becomes thermodynamically unstable at sufficiently low temperatures \cite{Anabalon:2024cnb}. Our results indicate that the natural resolution is for the black brane to acquire angular momentum. Rotation gives rise to a family of everywhere regular supersymmetric ground states, providing a natural endpoint for the decay of electrically charged configurations. Our solutions seem to supply the missing supersymmetric endpoint of the instability of electrically charged AdS black branes.

It has been known for some time that for spinning black holes temperature is not the black hole temperature and that the fluid entropy is not the black hole entropy density. They coincide only in the large-black-hole limit \cite{Bhattacharyya:2007vs}. Indeed, the CFT, for which the ABJM theory is the primer in 3d, has 
degrees of freedom with spin: spin 1/2 fermions, as well as CS
gauge fields. In a future paper we shall see that by adding this spin density to the Euler relation one can formulate a consistent thermodynamical description in terms of the black hole variables of this paper. 

Taken together, these results provide a new framework for the holographic description of strongly coupled quantum field theories in Minkowski spacetime and open several avenues for future investigation.
\section*{Acknowledgements}
The work of HN is supported in part by  CNPq grant 
304583/2023-5 and FAPESP grant 2024/15298-0.
HN would also like to thank the ICTP-SAIFR for their support 
through FAPESP grant 2021/14335-0. The work of AA is supported in part by the ANID FONDECYT grants 1230853, 1242043, 1250133, 1262452 and 1262414 and by the FAPESP grant 2024/16864-9. 

\newpage

\hypersetup{linkcolor=blue}
\phantomsection
\addtocontents{toc}{\protect\addvspace{4.5pt}}

\bibliographystyle{mybibstyle}
\bibliography{bibliografia}
\end{document}